\documentclass[a4paper,fleqn,usenatbib]{mnras}

\usepackage{mathptmx}

\usepackage[T1]{fontenc}
\usepackage{ae,aecompl}

\usepackage{graphicx}	
\usepackage{amsmath}	
\usepackage{amssymb}	

\title[signatures of a new-born black hole in short GRBs]{Search for the Signatures of a New-Born Black Hole from the Collapse of a Supra-massive Millisecond Magnetar in Short GRB Light Curves}

\author[Q. Zhang et al.]{
Q. Zhang,$^{1}$\thanks{E-mail: zhangqiang@ihep.ac.cn}
W. H. Lei,$^{2}$
B. B. Zhang$^{3,4}$
W. Chen,$^{2}$
S. L. Xiong,$^{1}$
and L. M. Song$^{1}$
\\
$^{1}$Key Laboratory of Particle Astrophysics, Institute of High Energy Physics, Chinese Academy of Sciences, Beijing 100049, China\\
$^{2}$School of Physics, Huazhong University of Science and Technology, Wuhan 430074, China\\
$^{3}$Instituto de Astrof\'{i}sica de Andaluc\'{a} (IAA-CSIC), P.O. Box 03004, E-18080 Granada, Spain\\
$^{4}$Scientist Support LLC, Madsion, AL 35758, USA}

\date{Accepted XXX. Received YYY; in original form ZZZ}

\pubyear{2017}

\begin{document}
\label{firstpage}
\pagerange{\pageref{firstpage}--\pageref{lastpage}}
\maketitle

\begin{abstract}
`Internal plateau' followed by a sharp decay  is commonly seen in short gamma-ray burst (GRB) light curves.
The plateau component is usually interpreted as the dipole emission from a supra-massive magnetar, and the sharp decay
may imply the collapse of the magnetar to a black hole (BH).
Fall-back accretion onto the new-born BH could produce long-lasting activities via the Blandford-Znajek (BZ) process.
The magnetic flux accumulated near the BH would be confined by the accretion disks for a period of time.
As the accretion rate decreases, the magnetic flux is strong enough to obstruct gas infall,
leading to a magnetically-arrested disk (MAD). Within this scenario,
we show that the BZ process could produce two types of typical X-ray light curves:
type I exhibits a long-lasting plateau, followed by a power-law decay
with slopes ranging from 5/3 to 40/9; type II shows roughly a single power-law decay with slope of 5/3.
The former requires low magnetic filed strength, while the latter corresponds to relatively high values.
We search for such signatures of the new-born BH from a sample of short GRBs with an
internal plateau, and find two candidates: GRB 101219A and GRB 160821B, corresponding to type II and type I light curve,
respectively. It is shown that our model can explain the data very well.
\end{abstract}



\section{INTRODUCTION}\label{intro}
Short gamma-ray bursts (GRBs), with durations typically less than 2~s \citep{kou93}, have been widely speculated to be produced by mergers of two compact objects:
either a double neutron star or a neutron star (NS) and a black hole (BH) binary \citep{eich89,pacz91,nara92}.
This is observationally supported by their host galaxy properties, the locations of the bursts in the host galaxies,
as well as non-detection of supernova associations \citep[e.g.,][]{bart05b,berg05,fox05,gehr05,fong10,kann11,berg14}.
While NS-BH mergers inevitably end up in a BH instantly, NS-NS mergers can result in different types of remnants.
Depending on the total mass of the NS-NS system and the NS equation of state (EOS), the final products of NS-NS mergers could be a prompt BH
\footnote{Here `prompt BH' means that the merger product may either immediately collapse to a BH or survive for $\sim$ 10--100 ms as a hypermassive neutron supported by strong differential rotation and thermal pressures \citep[e.g.,][]{baio08,kiuc09,rezz11,hoto13}.}, a supra-massive NS (SMNS) temporarily supported by uniform rotation and collapses to a BH until the centrifugal force is insufficient to support the mass, or a stable NS \citep[e.g.,][]{ross02,giac13,ravi14,ciol17}.

The discovery of $\sim 2M_{\sun}$ NSs suggests that the EOS is stiff enough for SMNSs to be created from the merger of two NSs \citep{demo10,anto13,hebe13}.
Moreover, observations of the early afterglows of a large sample of short GRBs with {\it Swift} show rich
features that indicate extended engine activities,
such as extended emission \citep{norr06}, X-ray flares \citep{bart05b,camp06,marg11} and X-ray plateaus \citep{rowl13,lv15}.
These features are hard to
interpret within the framework of a BH central engine, but are compatible with a rapidly spinning,
strongly magnetized NS or `millisecond magnetar' as
the central engine \citep[e.g.,][]{dai06,gao06,metz08b,rowl10,rowl13,gomp13,gomp14}.

A good fraction of short GRBs detected with {\it Swift} show `internal X-ray plateaus' within the first few hundred seconds, followed by a sharp drop with a temporal decay
index of $\alpha>3$\footnote{The convention $F_{\nu}\propto\nu^{-\beta}t^{-\alpha}$ is adopted throughout the paper, where $\beta$ is the spectral index and $\alpha$
is the temporal decay index.}, sometimes approaching $\alpha\sim 10$ \citep{rowl10,rowl13,lv15}. Such an internal plateau has also been observed in several long GRBs \citep{troj07,lyon10,lv14}, but they are commonly observed in short GRBs. This kind of plateau, followed by a rapid decay which is too steep to be explained within the external shock model,
can be interpreted as the internal emission of the magnetar wind, and the sharp decay marks the abrupt cessation of the central engine, likely due to
the collapse of a supra-massive magnetar to a BH \citep{troj07,rowl10,zhan13,zhan14}. Since the GRB outflow still produces X-ray afterglow by the external shock during the internal plateau phase, it is expected to emerge once the X-ray emission from the magnetar wind drops below the external component. This has been seen clearly in several X-ray afterglows of both long
and short GRBs \citep{troj07,lyon10,rowl13,lv14,lv15,de16,zhan16}. The external component typically shows a power-law (PL) decay with slope of $\sim1$ which is consistent with the prediction of the standard afterglow models \citep{sari98,chev00}.

Nevertheless, the recent observed GRB 160821B seems to challenge the above scenario. GRB160821B is a nearby bright short GRB
detected by {\it Swift} and {\it Fermi} \citep{sieg16,stan16}, with a redshift of $z=0.16$ \citep{leva16}. Its X-ray afterglow exhibits an internal plateau lasting for $\sim$ 180~s, then drops steeply with a slope of $\sim4.5$. About 1000~s after the Burst Alert Telescope \citep[BAT;][]{bart05a} trigger, another plateau component emerges, which
lasts for $\sim3\times10^4$~s with a decay slope of $\sim0.45$, then the light curve declines with a slope of $\sim3.5$ \citep{lv17}.
This `late plateau'\footnote{Hereafter we use the term `late plateau' to denote the long-lasting plateau following the internal plateau and the sharp decay phase.} is not expected within the standard afterglow models, and additional energy injection is needed.
If we assume the sharp decay following the internal plateau is  due to the collapse of a supra-massive magnetar, then the energy injection required by the late plateau of GRB 160821B can only be provided by the new-born BH. In this case, the late X-ray plateau suggests a possible signature of a new-born BH from the collapse of a supra-massive magnetar.

Motivated by the special X-ray light curve of GRB 160821B, in this paper, we attempt to answer the following questions:
What are the possible mechanisms for the long-lasting X-ray emission after the collapse of a supra-massive magnetar to a BH?
What types of X-ray signatures can be expected from the new-born BH? Besides GRB 160821B, are there other short GRBs with an internal plateau
showing these X-ray signatures? Recently, \citet{chen17} found that the X-ray bump following the internal plateau
of long GRB 070110 is a possible signature of a new-born BH from the collapse of a supra-massive magnetar and interpreted the X-ray bump as the result of a fall-back accretion onto the BH. Their work further encourages us to explore the possible X-ray signatures of the new-born BH in the
case of NS-NS mergers, and to search for these signatures in the sample of short GRBs that show an internal plateau.

This paper is organized as follows. In Section \ref{model}, we give a general picture of the spin-down of a supra-massive magnetar and the fall-back
accretion onto the new-born BH in the case of NS-NS mergers. In particular, we predict the possible X-ray signatures of the new-born BH
that could be observed in the afterglows of short GRBs. We then search for the candidates that might show the BH signatures in their X-ray light curves using the sample of short GRBs with an internal plateau and compare our model with observations in Section \ref{obs}. Finally, we briefly summarize our results and discuss the implications in Section \ref{conclu}. Throughout the paper, we use the standard notation $Q_x=Q/10^{x}$ with $Q$ being a generic quantity in cgs units, and a concordance cosmology with parameters $ H_0=71 ~\rm km \ s^{-1} Mpc^{-1}$, $\rm \Omega_M = 0.30$,
and $\Omega_\Lambda =0.70$ is adopted \citep{jaro11}. All the errors are quoted at the $1\sigma$ confidence level.

\section{model}\label{model}
Our model assumes that the product of the NS-NS merger is a supra-massive magnetar, as it spins down, it collapses to a BH when the
centrifugal force is insufficient to support the mass. We interpret the internal plateau of short GRBs as the magnetic dipole emission from the magnetar
and attribute the late X-ray emission (for GRB 160821B, it is a late plateau plus a steep decay) to a process
of fall-back accretion onto the new-born BH. Our main work in this section is to explore the possible long-lasting X-ray signatures
of the new-born BH.

\subsection{Dipole Spin-Down of a Supra-massive Magnetar} \label{spindown}
Numerical simulations show that NS-NS mergers could eject a fraction of the materials,
forming a mildly anisotropic outflow (the so-called `dynamical ejecta')
with a typical mass of $\sim10^{-4}-10^{-2}~M_{\sun}$ and a typical velocity of $\sim 0.1-0.3~c$ \citep[e.g.,][]{rezz11,hoto13,ross13,ciol17}.
The dynamical ejecta are followed by a slower outflow of material that does not exceed the escape velocity and might fall back onto the remnant
at later times. The amount of fall-back matter is comparable to or larger than that of the escaping ejecta \citep[e.g.,][]{ross07,hoto13,ciol17}, and this material is prone to return within a few seconds and create a new ring at a radius of around 300--500~km \citep{lee09}.

In the case of the millisecond magnetar engine, the magnetar would interact with the infalling material via accretion and propeller
processes \citep[e.g.,][]{piro11}. These processes affect the dipole spin-down and may produce intense electromagnetic emission
\citep[e.g.,][]{gomp14,gibs17}. However, considering the small fall-back mass combined with accretion disc heating effects, the influence on the
spin period is not important \citep{rowl13,ravi14}.
In addition, the post-merger magnetar may undergo important gravitational wave (GW) radiation \citep[e.g.,][]{zhan01,cors09,fan13,dall15,done15,lask16,gao17}, during which a significant spin energy is taken away by GWs.
This affects the magnetar spin-down and the collapse time \citep{gao16}. Alternatively, the GW effect could be effectively taken into account by choosing a relatively large initial spin period within the dipole spin-down framework as a first approximation \citep[e.g.,][]{rowl13,lv15}.
In fact, we will see in this subsection that one can give only the upper limits of the magnetar parameters (period and magnetic field strength)
by modeling the observed internal plateaus, considering the above effects would complicate our calculations and give no meaningful results.
In this work, we thus do not consider the effects of the magnetar accretion and propeller and GW losses on its spin evolution,
and use the simple dipole spin-down model \citep{zhan01}.

The characteristic spin-down luminosity $L_0$ and the characteristic spin-down timescale $\tau$  are
related to the magnetar initial parameters as \citep{zhan01}
\begin{eqnarray}\label{L0}
L_0 &=& 1.0\times 10^{49}~{\rm erg~s}^{-1}\left(B_{\rm p,15}^2 P_{0,-3}^{-4}R_6^6\right), \\ \label{tau}
\tau &=& 2.05\times10^3~{\rm s} \left(I_{45}B_{\rm p, 15}^{-2} P_{0,-3}^{2} R_{6}^{-6}\right),
\end{eqnarray}
where $I$ is the moment of inertia, $B$ is the surface magnetic field strength at the poles, $P_0$ is the initial spin period, and $R$ is the radius of the magnetar.

The isotropically equivalent luminosity of the internal plateau ($L_{\rm int}$) is related to the spin-down luminosity $L_0$ as
\begin{equation}
L_{\rm int}=\left(\eta_{\rm X}/f_{\rm b}\right)L_0, \label{L_int}
\end{equation}
where $\eta_{\rm X}$ is the radiation efficiency, and $f_{\rm b}=1-\cos \theta_{\rm j}$ is the beaming factor.

The spin-down formula due to dipole radiation is given by
\begin{equation}
P(t)=P_0\left(1+\frac{t}{\tau}\right)^{1/2}. \label{Pt}
\end{equation}
A supra-massive magnetar is temporarily supported by rigid rotation, which could enhance the maximum gravitational mass ($M_{\rm max}$)
allowed for NS surviving. For a given EOS, one can write $M_{\rm max}$ as a function of the spin period $P$ \citep{lyfo03},
\begin{equation}
M_{\rm max}=M_{\rm TOV}\left(1+\hat{\alpha} P^{\hat\beta}\right), \label{Mmax}
\end{equation}
where $M_{\rm TOV}$ is the maximum mass for a non-rotating NS, $\hat\alpha$ and $\hat\beta$ depend on the EOS.

The supra-massive magnetar collapses when its spin period becomes large enough that $M_{\rm max}(P)=M_{\rm p}$, where $M_{\rm p}$
is the mass of the protomagnetar. Using Equations (\ref{Pt}) and (\ref{Mmax}), one can derive the collapse time $t_{\rm col}$ \footnote{When the GW effect is considered, the expression of $t_{\rm col}$ is different from Equation (\ref{tcol}) and has been derived by \citet{gao16}.
We refer the reader to see this paper for details.}\citep{lask14,lv15}, i.e.,
\begin{eqnarray}
t_{\rm col}&=&\frac{3c^3I}{4\pi^2 B_{\rm p}^2 R^6}\left[\left(\frac{M_{\rm p}-M_{\rm TOV}}{\hat\alpha M_{\rm TOV}}\right)^{2/\hat\beta}-P_{0}^2\right]\nonumber\\
           &=&\frac{\tau}{P_0^2}\left[\left(\frac{M_{\rm p}-M_{\rm TOV}}{\hat\alpha M_{\rm TOV}}\right)^{2/\hat\beta}-P_{0}^2\right]. \label{tcol}
\end{eqnarray}

The collapse time of the supra-massive magnetar can be generally identified as the plateau break time in the source frame, i.e., $t_{\rm col}\simeq t_{\rm b,int}$.
Since the post-plateau decay slope is typical steeper than 3, the spin-down timescale should be greater than the break time. We thus take $t_{\rm col}$
as the lower limit of the spin-down timescale. The magnetar parameters $P_0$ and $B_{\rm p}$ can be solved from Equations (\ref{L0}) and (\ref{tau}), i.e.,
\begin{eqnarray}\label{P0}
P_{0,-3} &=& 1.42~{\rm s} \left(I_{45}^{1/2}L_{0,49}^{-1/2}\tau_{3}^{-1/2}\right), \\ \label{Bp}
B_{\rm p,15} &=& 2.05~G \left(I_{45} R_6^{-3} L_{0,49}^{-1/2}\tau_3^{-1}\right).
\end{eqnarray}
With the plateau luminosity $L_{\rm int}$ and the break time $t_{\rm b,int}$, one can derive the upper limits
of $P_0$ and $B_{\rm p}$ from Equations (\ref{L_int}), (\ref{P0}) and (\ref{Bp}) when the NS EOS and the value of $\eta_{\rm X}/f_{\rm b}$ are assumed.
When a reasonable value of $P_0$ in the range of $P_{0,\rm min}\leqslant P_0 \leqslant P_{0,\rm max}$ is
adopted\footnote{Here $P_{0,\rm min}$ is the break-up limit and $P_{0,\rm max}$ is the derived upper limit of the spin period.}, we can derive $M_{\rm p}$ from Equation (\ref{tcol}) based on the data and a given EOS.

The NS EOS is most uncertain. Using the general relativistic hydrostatic equilibrium code RNS \citep{ster95}, the numerical values for $M_{\rm TOV}$, $R$, $I$ and
thus $\hat\alpha$ and $\hat\beta$ for several EOSs have been worked out \citep{lask14}. More EOSs, especially those for quark star, were studied in \citet{li16}.
In this work, we adopt the EOS GM1 ($M_{\rm TOV}=2.37~M_{\sun}$, $R=12.05~{\rm km}$,
$I=3.33\times10^{45}~{\rm g~cm}^{-2}$, $\hat\alpha=1.58\times10^{-10}{\rm s}^{-\hat\beta}$ and $\hat\beta=-2.84$), which is favored by the short GBR data
under the assumption that the cosmological NS-NS merger systems have the same mass distribution as the observed Galactic NS-NS population \citep{lask14,lv15,gao16}.

\subsection{Fall-Back Accretion onto the New-Born BH} \label{fallback}

Fall-back accretion onto a central BH and the resulting radiation have been intensively studied in the framework of a prompt BH \citep[e.g.,][]{ross07,metz08a,lee09,rossi09}. After the original accretion discs are consumed on a viscous timescale of $\sim0.1$~s,
the BH begins to accrete the fall-back matter, the accretion rate of which follows a single PL,
i.e., $\dot M_{\rm fb}\propto t^{-5/3}$ \citep[e.g.,][]{ross07}.

If the BH is produced by the collapse of a supra-massive magnetar as considered in this work, there is no debris disk left \citep{marg15}.
Before the BH forms, the fall-back material has already returned and created a disk at a radius of a few hundred kilometres.
The magnetar accretion and propeller processes would inevitably decrease the total fall-back mass left for the BH to accrete.
We thus expect a smaller fall-back mass $M_{\rm fb}$ in our magnetar scenario than the results obtained from the numerical simulations \citep[e.g.,][]{ross07,hoto13,ciol17}.

Another issue to be specified is the mass accretion rate ($\dot M$) of the new-born BH.
The relation between $\dot M$ and $\dot M_{\rm fb}$ is uncertain,
but it is plausible to assume that $\dot M$ is a fraction of $\dot M_{\rm fb}$\footnote{For the mass accretion rate of a BH, \citet{metz08a} found $\dot M\propto t^{-4/3}$ for an advection-dominated disk, while the numerical simulation results of \citet{fern15}
showed $\dot M\propto t^{-2.2}~(1~{\rm s}\lesssim t\lesssim 10~{\rm s})$
when the display between the disk winds and the dynamical ejecta is considered.
However, it is unclear whether these scale laws apply to large time scales (e.g., $t\gtrsim100$~s).} \citep[e.g.,][]{tche14,kisa15}, i.e.,
\begin{equation}
\dot M=f_{\rm acc} \dot M_{\rm fb}=\dot M_{\rm i}\left(\frac{\tilde t}{\tilde {t}_{\rm PL}}\right)^{-5/3},~~(\tilde t>\tilde {t}_{\rm PL}), \label{M_dot}
\end{equation}
where $0<f_{\rm acc}\leqslant 1$ is a proportionality constant, $\dot M_{\rm i}=2 f_{\rm acc} M_{\rm fb}/(3 \tilde {t}_{\rm PL})$ is
the initial mass accretion rate, $\tilde{t}$ is the time since the BH accretion, and $\tilde t_{\rm PL}$ denotes the beginning time of such a
PL accretion. We use $t_0$ to denote the beginning time of the BH accretion
and approximately set it to be the collapse time, i.e., $t_0\simeq t_{\rm col}$.
The value of $\tilde {t}_{\rm PL}$ is uncertain, but a conservative estimation of $\tilde {t}_{\rm PL}\sim 1$~s seems to be reasonable \citep[e.g.,][]{ross07,metz08a,lee09,fern15}.

Energy extraction from the BH-accretion disk system can be via neutrino-antineutrino annihilation \citep{poph99,nara01,dima02,gu06,chen07,jani07,lei09,liu15,xie16},
or Blandford-Znajek mechanism \citep[BZ, hereafter;][]{blan77,lee00,li00,lei05,lei13}.
In the case of fall-back accretion, the neutrino-antineutrino annihilation becomes inefficient quickly \citep[$>0.1$~s;][]{rossi09,metz08a}
and cannot explain the late X-ray emission observed in short GRBs with an internal plateau (e.g., the late plateau of GRB 160821B).
The BZ process remains a possible mechanism that power the long-lasting X-ray emission.

The BZ process extracts the BH rotational energy via the large-scale poloidal magnetic field that is supported by the surrounding torus \citep{blan77,lee00}.
For a Kerr BH with mass $M_{\bullet}(\equiv m_{\bullet} M_{\sun})$ and angular momentum $J_{\bullet}$, the BZ power can be estimated as \citep{lee00,li00,wang02,mcki05,lei11,lei13}
\begin{equation}
L_{\rm BZ}= 1.7\times 10^{50}~{\rm erg~s}^{-1}\ a_{\bullet}^2 m_{\bullet}^2 B_{\bullet,15}^2 F(a_{\bullet}), \label{LBZ}
\end{equation}
where $a_{\bullet}=J_{\bullet} c/\left(G M_{\bullet}^2\right)$ is the dimensionless spin parameter of the BH, $B_{\bullet}$ is the magnetic field
strength threading the BH horizon.
The spin-dependent function $F(a_{\bullet})$ can be approximated as \citep{lee00,wang02}
\begin{equation}
F(a_{\bullet})=\left[\left(1+q^2\right)/q^2\right]\left[\left(q+1/q\right) \arctan q-1\right], \label{Fa}
\end{equation}
where $q=a_{\bullet}/(1+\sqrt{1-a_{\bullet}^2})$, and $2/3 \leqslant F(a_{\bullet}) \leqslant \pi-2$ for $0\leqslant a_{\bullet} \leqslant 1$.
\citet{tche10} investigated this function numerically and gave an analytical fit to the numerical model. Their obtained function is similar to
Equation (\ref{Fa}) at most $a_{\bullet}$ values and only slightly deviates from it as $a_{\bullet}$ approaches 1. We thus use Equation (\ref{Fa})
as a reasonable approximation.

A major uncertainty in estimating the BZ power is the strength of magnetic fields \citep{kuma15}. Because of the freedom in $B_{\bullet}$, for a fixed value of BH spin
$a_{\bullet}$ and mass accretion rate $\dot M$, we expect a range of BZ powers from zero (no jet) up to a maximum value \citep{tche15book}.
By performing advanced numerical simulations,  \citet{tche11} demonstrated that accretion disks can accumulate large-scale magnetic flux
($\Phi_{\bullet}\sim\pi r_{\bullet}^2 B_{\bullet}$) on the BH,
until the magnetic flux becomes so strong that it obstructs gas infall and leads to a magnetically-arrested disk \citep[MAD;][]{bisn74,bisn76,igum03,nara03}.
Since the BH magnetic flux is maximum in the MAD state, MADs achieve the maximum possible efficiency of jet production\citep{tche11}.
These results were later used in the areas of active galactic nuclei, tidal disruptions events, and GRBs \citep[e.g.,][]{tche14,zama14,kisa15,tche15}.

The MAD state happens at a critical mass accretion rate that can be estimated by assuming the magnetic pressure ($P_{\rm mag}$) and the disk gas pressure ($P_{\rm gas}$)
balance at the BH horizon \citep[e.g.,][]{nara03,kisa15,tche15book}, i.e.,
\begin{equation}
\frac{B_{\bullet}^2}{8\pi}=\frac{GM_{\bullet} \dot M}{2\pi r^3 \upsilon_{r}}\Big |_{r=r_{\bullet}}, \label{balance}
\end{equation}
where $G$ is the gravitational constant, $\dot M$ is the mass accretion rate of the BH, $r_{\bullet}=(1+\sqrt{1-a_{\bullet}^2})r_{\rm g}\equiv \chi(a_{\bullet})r_{\rm g}$
is the radius of the BH horizon, and $r_{\rm g}=G M_{\bullet} /c^2$;
$\upsilon_r=\epsilon \upsilon_{\rm ff}$
is the radial velocity of the infalling gas outside the horizon, where $\upsilon_{\rm ff}=\sqrt{GM_{\bullet}/r}$ is the free-fall velocity.
Since the accretion disk gas diffuses towards the BH via magnetic reconnection and interchanges, the velocity is much less than the free-fall velocity.
It is reasonable to adopt $\epsilon=10^{-2}$ \citep{nara03,kisa15}, which is supported by the observations and numerical simulations of
the relativistic jets \citep[e.g.,][]{tche11,zama14}. From Equation (\ref{balance}), one can derive
\begin{eqnarray}
\dot M_{\rm MAD}&=&8.2\times 10^{-12}~M_{\sun}~{\rm s}^{-1}\ \chi^{5/2}\left(\frac{\epsilon}{10^{-2}}\right) \nonumber \\
&  & \times\left(\frac{B_{\bullet}}{10^{12}~{\rm G}}\right)^2 \left(\frac{m_{\bullet}}{3}\right)^2.  \label{dot_M_MAD}
\end{eqnarray}

The initial accretion rate
$\dot M_{\rm i}=2f_{\rm acc}M_{\rm fb}/(3 \tilde {t}_{\rm PL})\sim 10^{-5}-10^{-3} M_{\sun}~{\rm s}^{-1}$
if we adopt $M_{\rm fb}\sim 10^{-4}-10^{-2} M_{\sun}$, $f_{\rm acc}=0.5$ and $\tilde {t}_{\rm PL}=1$~s.
As long as $B_{\bullet}$ is not much larger than $10^{14}$~G, we have $\dot M_{\rm i}\gg \dot M_{\rm MAD}$.
According to Equation (\ref{balance}), this is equivalent to say $P_{\rm gas}\gg P_{\rm mag}$.
In this case, the gas accreted onto the new-born BH is more than sufficient to confine the magnetic flux within the BH vicinity.
As long as $\dot M\geqslant\dot M_{\rm MAD}$,
the BZ power is determined by the magnetic flux $\Phi_{\bullet}$ and not by mass accretion rate. During this phase (`pre-MAD' hereafter), the BZ power
$L_{\rm BZ}\propto a_{\bullet}^2 m_{\bullet}^2 B_{\bullet}^2 \propto a_{\bullet}^2 m_{\bullet}^{-2}\Phi_{\bullet}^2$ is roughly a constant \citep{tche15,kisa15}.
As $\dot M$ deceases and eventually drops below $\dot M_{\rm MAD}$, we have $P_{\rm gas}<P_{\rm mag}$, then the magnetic flux becomes dynamically important,
and parts of the flux diffuse out. The remaining magnetic flux obstructs gas infall and leads to a MAD \citep{tche15}.
In the MAD regime, the magnetic field (and thus the magnetic flux) is determined by the instantaneous $\dot M$ via Equation (\ref{dot_M_MAD}), i.e.,
$B_{\bullet}\propto \dot M^{1/2}$ \citep{tche15}. Thus, the BZ power $L_{\rm BZ}\propto B_{\bullet}^2\propto \dot M$ is a function of mass accretion rate.

Taking into account the X-ray radiation efficiency $\eta_{\bullet, \rm X}$ and the beaming factor $f_{\bullet, \rm b}$,
the observed isotropic X-ray luminosity can be written as
\begin{equation}
L_{\rm X, iso}=(\eta_{\bullet, \rm X}/f_{\bullet, \rm b})L_{\rm BZ}. \label{LxfromBZ}
\end{equation}

We assume that the MAD is achieved at $\tilde{t}=\tilde{t}_{\rm MAD}$. In this time coordinate system, the light curve shows a plateau
in the pre-MAD regime ($\tilde{t}\leqslant\tilde{t}_{\rm MAD}$), followed by a single PL with the decay slope consistent with $\dot M$ in the MAD regime
($\tilde{t}>\tilde{t}_{\rm MAD}$).

The plateau luminosity $L_{\rm preMAD}$ can be obtained straightforwardly
from Equations (\ref{LBZ}) and (\ref{LxfromBZ}),
\begin{eqnarray}
L_{\rm preMAD}&=&1.5\times10^{43}~{\rm erg~s}^{-1}\ \left(\frac{\eta_{\bullet, \rm X}/f_{\bullet, \rm b}}{1}\right)\left(\frac{a_{\bullet}}{0.1}\right)^2  \nonumber \\
               &  & \times F(a_{\bullet})\left(\frac{B_{\bullet,\rm MAD}}{10^{12}~{\rm G}}\right)^2 \left(\frac{m_{\bullet}}{3}\right)^2, \label{L_lp}
\end{eqnarray}
where $B_{\bullet, \rm MAD}$ is the critical magnetic field strength that is related to $\dot M_{\rm MAD}$ by Equation (\ref{dot_M_MAD}).
It is also the magnetic field strength in the pre-MAD regime and can be determined by observations.
The duration of the pre-MAD state and thus the break time of the X-ray plateau can be derived
from $\dot M(\tilde{t}_{\rm MAD})=\dot M_{\rm MAD}$.
Using Equations (\ref{M_dot}) and (\ref{dot_M_MAD}), we obtain
\begin{eqnarray}
\tilde{t}_{\rm MAD}&=&3.7\times10^4~{\rm s}\ \chi^{-3/2}\left(\frac{\epsilon}{10^{-2}}\right)^{-3/5}
\left(\frac{f_{\rm acc}}{0.5}\right)^{3/5} \nonumber \\
             &  & \times \left(\frac{\tilde {t}_{\rm PL}}{1~{\rm s}}\right)^{2/5} \left(\frac{m_{\bullet}}{3}\right)^{-6/5}\nonumber \\
            &  & \times\left(\frac{M_{\rm fb}}{10^{-3}M_{\sun}}\right)^{3/5} \times \left(\frac{B_{\bullet,\rm MAD}}{10^{12}~{\rm G}}\right)^{-6/5}. \label{t_blp}
\end{eqnarray}

During the MAD phase, the X-ray luminosity $L_{\rm X,iso}\propto \dot M\propto t^{-5/3}$.
Then we can model the luminosity evolution as
\begin{eqnarray}
L_{\rm X,iso}(\tilde{t})&=& L_{\rm preMAD}\left(1+\frac{\tilde{t}}{\tilde{t}_{\rm MAD}}\right)^{-5/3}, \nonumber\\
                &\approx & \left\{
                \begin{array}{ll} \label{Lxlate1}
                L_{\rm preMAD},  &  0<\tilde{t}\ll \tilde{t}_{\rm MAD},\\
                L_{\rm preMAD}\left(\frac{\tilde{t}}{\tilde{t}_{\rm MAD}}\right)^{-5/3},  & \tilde{t}\gg \tilde{t}_{\rm MAD}.
                \end{array}
                \right.
\end{eqnarray}
As seen from Equation (\ref{Lxlate1}), in the $\tilde{t}$ coordinate system, the predicted X-ray luminosity exhibits an initial plateau before $\tilde{t}_{\rm MAD}$, then decays as $\tilde{t}^{-5/3}$.

According to Equations (\ref{L_lp}) and (\ref{t_blp}), when the values of $a_{\bullet}$, $m_{\bullet}$, $B_{\bullet,\rm MAD}$ and $M_{\rm fb}$ are given, we can model the observed X-ray
light curve straightforwardly. Conversely, these parameters can be obtained by modeling the afterglow data. Among of them,  $B_{\bullet,\rm MAD}$ and $M_{\rm fb}$ can be
derived as follows.

From Equation (\ref{L_lp}), the magnetic field strength $B_{\bullet,\rm MAD}$ is given by
\begin{eqnarray}
B_{\bullet,\rm MAD} &=& 7.8\times10^{11}~{\rm G}\ \left(\frac{a_{\bullet}}{0.1}\right)^{-1}\left(\frac{m_{\bullet}}{3}\right)^{-1} F^{-1/2}(a_{\bullet})\nonumber \\
          & &\times\left(\frac{\eta_{\bullet, \rm X}/f_{\bullet, \rm b}}{1}\right)^{-1/2}
\left(\frac{L_{\rm preMAD}}{10^{43} ~{\rm erg~s}^{-1}}\right)^{1/2}. \label{B_preMAD}
\end{eqnarray}
Then the total fall-back mass $M_{\rm fb}$ can be obtained from Equations (\ref{t_blp}) and (\ref{B_preMAD}),
\begin{eqnarray}
M_{\rm fb} &=& 6.7\times10^{-5} M_{\sun}\ \left(\frac{f_{\rm acc}}{0.5}\right)^{-1}
\left(\frac{\eta_{\bullet,\rm X}/f_{\bullet,\rm b}}{1}\right)^{-1} \nonumber \\
          &  & \times \left(\frac{\epsilon}{10^{-2}}\right)\left(\frac{a_{\bullet}}{0.1}\right)^{-2}\chi^{5/2}F^{-1}(a_{\bullet}) \nonumber \\
          &  & \times \left(\frac{\tilde {t}_{\rm PL}}{1~{\rm s}}\right)^{-2/3}\left(\frac{L_{\rm preMAD}}{10^{43}~{\rm erg~s}^{-1}}\right)
\left(\frac{\tilde{t}_{\rm MAD}}{10^4~{\rm s}}\right)^{5/3}.  \label{Mfb}
\end{eqnarray}
We note that $M_{\rm fb}\propto \tilde {t}_{\rm PL}^{-2/3}$ is not strongly dependent on $\tilde {t}_{\rm PL}$,
so the value of $\tilde {t}_{\rm PL}$ adopted in our modeling will not have a significant influence on the derived $M_{\rm fb}$.

\begin{figure}
\includegraphics[width=\columnwidth]{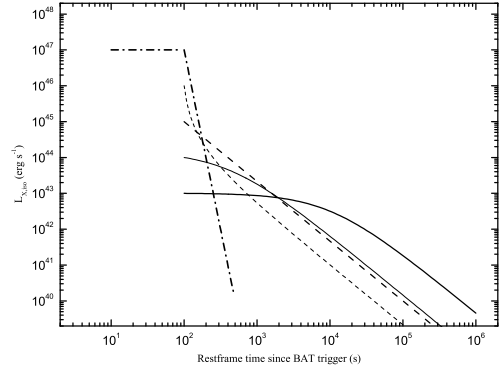}
\caption{Theoretical X-ray light curves produced by the dipole spin-down of a supra-massive magnetar and the BZ process of the new-born BH.
The magnetar radiation component is shown with dotted dash lines. The following magnetar parameters are adopted:
the spin-down luminosity $L_0=10^{47}~{\rm erg~s}^{-1}$, the collapse time $t_{\rm col}=100$~s and the decay slope following the collapse $\alpha=10$.
The BH radiation component is drawn according to Equation (\ref{Lxlate}).
A series of values of $L_{\rm preMAD}$ and $\tilde{t}_{\rm MAD}$
are used in order to show different luminosity evolutions. The adopted parameters and the resulting light curves are:
$L_{\rm preMAD}=10^{43}~{\rm erg~s}^{-1}$, $\tilde{t}_{\rm MAD}=10^4$~s (thick solid line); $L_{\rm preMAD}=10^{44}~{\rm erg~s}^{-1}$, $\tilde{t}_{\rm MAD}=500$~s (thin solid line);
$L_{\rm preMAD}=10^{45}~{\rm erg~s}^{-1}$, $\tilde{t}_{\rm MAD}=100$~s (thick dashed line); $L_{\rm preMAD}=10^{46}~{\rm erg~s}^{-1}$ and $\tilde{t}_{\rm MAD}=10$~s (thin dashed line).
\label{LC}}
\end{figure}

To compare with observations, we need to transform the coordinate system from $\tilde{t}$ to $t$ by doing $\tilde{t}\rightarrow t-t_{\rm col}$. Then the X-ray luminosity can be written as
\begin{equation}
L_{\rm X,iso}(t)= L_{\rm preMAD}\left(1+\frac{t-t_{\rm col}}{\tilde{t}_{\rm MAD}}\right)^{-5/3}, ~~~~t>t_{\rm col}.  \label{Lxlate}
\end{equation}
A plateau-like feature can be found only if $\tilde{t}_{\rm MAD}\gg t_{\rm col}$.
To show this clearly, we draw the theoretical light curves according to Equation (\ref{Lxlate}), with a series of values of $L_{\rm preMAD}$ from $10^{43}$ to $10^{46}~{\rm erg}~{\rm s}^{-1}$ and the corresponding $\tilde{t}_{\rm MAD}$ from $10^4$ to 10~s.
As seen from Figure \ref{LC}, the light curve pattern strongly depends on the value of $\tilde{t}_{\rm MAD}$. When $\tilde{t}_{\rm MAD}\gg t_{\rm col}$,
the light curve shows a long-lasting plateau (e.g., the thick solid line); it reveals a gradual transition from a plateau to a single PL as $\tilde{t}_{\rm MAD}$ approaches
$\sim t_{\rm col}$ (e.g., from the thin solid line to the thick dashed line); as $\tilde{t}_{\rm MAD}$ further decreases, the light curve shows an initial steep decay followed by a smooth transition to a single PL with slope of 5/3 (e.g., the thin dashed line). We note that the light curve pattern is very sensitive to the values of
$L_{\rm preMAD}$ and $\tilde{t}_{\rm MAD}$. Thus, these parameters can, in principle, be obtained by modeling the afterglow data with Equation (\ref{Lxlate}).

The physics is easy to understand. With Equations (\ref{LBZ}), (\ref{dot_M_MAD}) and
$\dot M_{\rm MAD}=\dot M_{\rm i}(\tilde{t}_{\rm MAD}/\tilde {t}_{\rm PL})^{-5/3}$,
we have $\tilde{t}_{\rm MAD}\propto B_{\bullet, \rm MAD}^{-6/5}\dot M_{\rm i}^{3/5}$ and $L_{\rm preMAD}\propto B_{\bullet, \rm MAD}^2$.
For the same $\dot M_{\rm i}$, higher $B_{\bullet, \rm MAD}$ corresponds to smaller $\tilde{t}_{\rm MAD}$ and higher $L_{\rm preMAD}$, and
vice versa. Therefore, if the pre-MAD state can sustain a high magnetic field strength, say, $B_{\bullet,\rm MAD}\sim 10^{14}$~G, it would result in a small $\tilde{t}_{\rm MAD}\sim t_{\rm col}$ according to Equation (\ref{t_blp}) for typical parameters.
Though the light curve shows a short plateau in the $\tilde{t}$ coordinate system,
this plateau cannot be seen after the time transformation. In this case, the light curve is roughly a single PL.
Conversely, if the pre-MAD state has a weak magnetic field, say, $B_{\bullet,\rm MAD}\sim 10^{12}$~G, then the resulting $\tilde{t}_{\rm MAD}$ is
much longer than $t_{\rm col}$. This produces a long-lasting plateau in the light curve, followed by a temporal decline with slope of 5/3.
Therefore, the light curve patterns are mostly determined by the pre-MAD magnetic field strength. They are, however, also weakly affected by the initial accretion rate or the total fall-back mass
via $\tilde{t}_{\rm MAD}\propto B_{\bullet, \rm MAD}^{-6/5}\dot M_{\rm i}^{3/5}\propto B_{\bullet, \rm MAD}^{-6/5}M_{\rm fb}^{3/5}$.

It should be noted that, when the pre-MAD phase lasts a very long time, say, $\tilde{t}_{\rm MAD}\sim 10^4$~s,
the critical accretion rate is $\dot M_{\rm MAD}\sim 10^{-11}M_{\sun}~{\rm s}^{-1}$ according to Equation (\ref{dot_M_MAD}). Whether such a low accretion rate can sustain
the MAD state is questionable. \citet{kisa15} considered an extreme case by assuming that the fall-back matter is too small to support the magnetic flux at the
end of the pre-MAD regime. In this case, the MAD state cannot be achieved, and the magnetic field lines escape rapidly from the BH. They thus used a very large
post-plateau decay slope of 40/9\footnote{This value was derived based on the flux conversation.
The balance between the magnetic pressure and the gas pressure
gives the magnetospheric radius $r_{\rm m}$. As $\dot M$ deceases, $r_{\rm m}$ expands as $r_{\rm m}\propto t^{10/9}$. The magnetic flux $\Phi_{\bullet}\propto r_{\rm m}^{-2}$,
then $L_{\rm X}\propto L_{\rm BZ}\propto \Phi_{\bullet}^2\propto t^{-40/9}$ \citep{kisa15}.}.
For a short pre-MAD, however, $\dot M_{\rm MAD}$ is relatively high and we assume the MAD state can be achieved\footnote{In principle,
we should expect a change of slope from 5/3 to 40/9 at late times when the MAD state cannot be sustained. However, since it is difficult to predict when such a change occurs, we do not consider this extreme case and simply assume that the MAD state lasts long enough.}.
Therefore, we use the decay index of $\alpha_{\rm I}=5/3$ to 40/9 in Equation (\ref{Lxlate}) only for the light curve that shows a long-lasting plateau.
For other cases, we use $\alpha_{\rm II}=5/3$  that is consistent with the decay of $\dot M$ in the MAD state.

In summary, when the entire physical processes from a supra-massive magnetar to the new-born BH are considered,
our model predicts two types of typical X-ray light curves (see Figure \ref{LC} for the detailed evolutions however):
\begin{itemize}
\item Type I: an internal plateau plus a sharp drop with slope of $\alpha>3$, followed by a long-lasting plateau plus a steep decay with
slope of $\alpha_{\rm I}=5/3$ to 40/9. The luminosity of the late plateau is roughly constant and the duration is $\sim10^4$~s for typical parameters (see Equation (\ref{t_blp}));
\item Type II: an internal plateau plus a sharp drop, followed by a PL decay with slope of $\alpha_{\rm II}=5/3$.
\end{itemize}

We emphasize that type I and II light curves are intrinsically produced by the same physical process. It is the `zero point effect' that
makes their light curves look different. In the $\tilde{t}$ coordinate system, a plateau-like feature with duration $\sim \tilde{t}_{\rm MAD}$ always exists. When transforming to the $t$ coordinate system (the zero time is usually set to the trigger time), the existence of plateau
depends on the relation between the values of $t_0~(=t_{\rm col})$ and $\tilde{t}_{\rm MAD}$. Type I corresponds to $\tilde{t}_{\rm MAD}\gg t_0$ while type II requires that $\tilde{t}_{\rm MAD}$ is smaller or comparable with $t_0$.

\section{observational support}\label{obs}
In this section, we first search for the theoretically predicted X-ray light curves from a sample of short GRBs with an internal plateau, then compare our model with the data.
\subsection{Candidate Search}\label{search}
For a complete search of the candidates, we set up three criteria for sample selection: (1) We focus on short GRBs with an internal plateau that has
a post-plateau decay slope of $\alpha>3$. (2) We focus on bursts with high-quality post-plateau data, in particular, we
require the late X-ray data clearly show a feature deviating from the sharp decay phase, and span a wide range of time to show a clear temporal evolution.
(3) We further require the data to resemble our theoretically predicted light curves,  that is, the late X-ray data should
show a plateau with slope of $\sim 0$ or a single PL decay with slope of $\sim5/3$.

The properties of X-ray afterglow of short GRBs with an internal plateau have been systematically studied by \citet{rowl13}\footnote{The definition of `internal plateau'
used in our paper is actually consistent with the description in \citet{rowl13} that the X-ray plateau followed by a sharp drop which may suggest a magnetar collapsing to a BH.}
and \citet{lv15}. Using their samples, we find 11 firm candidates that clearly show an internal plateau. These candidates are the same as those bursts that were
marked with `unstable' magnetars in Table 6 of \citet{rowl13}.  This sample is further reduced based on criterion (2) and we find only two bursts that satisfy this requirement.
Together with GRB 160821B, the sample thus includes three bursts, which are GRB 070724A, GRB 101219A and GRB 160821B. As stated in Section \ref{intro}, GRB 160821B shows a
late plateau with slope of $\sim0.45$ and is marginally consistent with our type I light curve. The late X-ray data of GRB 101219A exhibit a single PL decay
with slope of $\sim 1.9$ \citep{evan09}, resembling our type II light curve. The light curve of GRB 070724A, however, is quite different. The late X-ray data show a
shallow decay with slope of $0.65^{+0.10}_{-0.12}$, followed by a steep decay with slope of $3^{+2}_{-1}$ \citep{ziae07}. The spectral index during this phase is
$\beta_{\rm X}\approx0.5$ \citep{koce10}. The slope of the shallow decay phase is consistent with the prediction of the standard afterglow model
when the X-ray frequency $\nu_{\rm X}$ is in the energy range of $\nu_{\rm m}<\nu_{\rm X}<\nu_{\rm c}$, where $\nu_{\rm m}$ and $\nu_{\rm c}$ are the typical synchrotron frequency and the cooling frequency of electrons, respectively \citep{sari98}.
This is also supported by the near-infrared (NIR) and optical afterglow observations \citep{berg09}. We thus do not consider this burst in the following analysis. Therefore, the final candidates include two bursts: GRB 101219A and GRB 160821B.

The {\it Swift} BAT and X-ray Telescope \citep[XRT;][]{burr05} data are downloaded from the {\it Swift} website\footnote{\url{http://www.swift.ac.uk/burst_analyser/}}.
The 0.3--10 keV unabsorbed X-ray flux was reduced by an automatic analysis procedure, and the BAT (15--150 keV) data were extrapolated to the
XRT band (0.3--10 keV) \citep{evan07,evan09}. Fortunately, both bursts have redshift measurements, we thus transform the flux data to
the luminosity light curve in the observed 0.3--10 keV energy band (see Figures \ref{160821B} and \ref{101219A}).

\subsection{Case Study}\label{case}
\subsubsection{GRB 160821B}
GRB 160821B triggered the {\it Swift}/BAT at 22:29:13 UT on 2016 August 21 \citep{sieg16}. It was
also detected by the {\it Fermi} Gamma-ray Burst Monitor (GBM) almost simultaneously \citep{stan16}. The BAT light curve shows a single short peak with
duration $T_{90}=0.48\pm0.07$~s \citep{palm16}. The time-integrated BAT+GBM spectrum can be jointly fit by a single PL with photon index
$\Gamma_{\gamma}=1.88\pm0.12$. The total fluence in the 8--1000 keV range is $(2.52\pm0.19)\times 10^{-6}$ erg cm$^{-2}$, with a redshift of
$z=0.16$ \citep{leva16}, this corresponds to an isotropically  equivalent energy
$E_{\gamma,\rm iso}=(2.1\pm0.2)\times 10^{50}$ erg \citep{lv17}.

The XRT began observing the field 66 s after the BAT trigger \citep{sieg16}.
The X-ray spectrum in the 0.3--10 keV energy band is best fit by an absorbed PL with photon index $\Gamma_{\rm X}=1.95^{+0.21}_{-0.08}$ and
column density $N_{\rm H}=(7.5\pm2.1)\times10^{20}$~cm$^{-2}$ \citep{lv17}. The light curve
shows an initial plateau lasting for $\sim180$~s then drops smoothly along with the $\Gamma_{\rm X}$ evolution
from $\sim 2$ to $\sim 3$. After about 1000~s, the light curve shows a late plateau followed by a steep decay and the photon index during this phase is $\sim 3$ \citep{lv17}.
The Ultra-Violet Optical Telescope \citep[UVOT;][]{romi05} began settled observations of the field of
GRB 160821B 76~s after the BAT trigger, but no optical afterglow consistent with the XRT position \citep{evan16} was detected.
Only preliminary 3$\sigma$ upper limits are obtained by using the UVOT photometric system for the first finding chart exposure \citep{bree16}.
In addition, possible macronova emission was reported in this burst \citep{troj16,kasl17}.

With the X-ray data of GRB 160821B and our model described in Section \ref{model}, we can now constrain the model parameters
($P_0$, $B_{\rm p}$, $M_{\rm p}$, $a_\bullet$, $B_{\bullet,\rm MAD}$, $M_{\rm fb}$) and then compare our model with the data.
For the magnetar parameters, we use the data of the internal plateau: $L_{\rm int}\simeq 1.2\times 10^{47}$~erg~s$^{-1}$ and
$t_{\rm b, int}\simeq180/(1+z)~{\rm s}=155~{\rm s}$ \citep{lv17}. By assuming $\eta_{\rm X}/f_{\rm b}=1$\footnote{The radiation efficiency $\eta_{\rm X}$
and the beaming factor $f_{\rm b}$ are unknown due to lack of knowledge on the jet production and dissipation process. Without loss of generality, we assume
$\eta_{\rm X}/f_{\rm b}=1$, which was used in \citet{rowl13} and \citet{lv15}. The statement above also applies to the new-born BH.
We thus adopt $\eta_{\bullet, X}/f_{\bullet,b}=1$ as well in the following estimations.}
and the EOS GM1, we obtain the upper limits of $P_0$ and $B_{\rm p}$ using Equations (\ref{L_int}), (\ref{P0}) and (\ref{Bp}),
i.e., $P_0\lesssim60$~ms, $B_{\rm p}\lesssim2.3\times10^{17}$~G. Within the magnetar model, \citet{rowl13} and \citet{lv15} investigated a dozen of short GRBs
with an internal plateau, and found that most bursts have $P_0$ and $B_{\rm p}$ values in the ranges
of 1--10~ms and $10^{15}$--$10^{16}$~G, respectively \citep[see also][]{gao16}. Without loss of generality, we adopt $P_0=4$~ms. As stated in Section \ref{model}, this relatively large $P_0$ is taken by equivalently considering angular momentum loss via strong GW radiation.
Then we have
$B_{\rm p}=1.0\times10^{15}$~G and $\tau=3.6\times10^4$~s according to Equations (\ref{L0}) and (\ref{tau}).
Using Equation (\ref{tcol}), we obtain the mass of the supra-massive magnetar $M_{\rm p}\simeq M_{\rm TOV}=2.37M_{\sun}$.

With the above magnetar parameters, we get the mass and spin of the new-born BH, i.e., $M_{\bullet}\simeq2.37M_{\sun}$ and $a_{\bullet}\simeq0.1$ (by using $J_{\bullet}=2\pi I/P_0$).
For other parameters, we use the data of the late plateau. Since the theoretically predicted post-plateau decay slope is uncertain, ranging from 5/3 to 40/9, we consider
two cases: (a) For $\alpha_{\rm I}=5/3$, we adopt the plateau luminosity
$L_{\rm preMAD}=8\times10^{43}$~erg s$^{-1}$ and the plateau duration $\tilde{t}_{\rm MAD}=3\times10^4/(1+z)~{\rm s}\simeq2.6\times10^4$~s.
By assuming $\eta_{\bullet, \rm X}/f_{\bullet, \rm b}=1$ and $f_{\rm acc}=0.5$, we obtain the magnetic field strength $B_{\bullet,\rm MAD}\simeq 3.1\times 10^{12}$~G and
the total fall-back mass $M_{\rm fb}\simeq 0.02 M_{\sun}$ from Equations (\ref{B_preMAD}) and (\ref{Mfb}).
We note that the value of $M_{\rm fb}$ is consistent with the obtained ejecta mass from
numerical simulations \citep[e.g.,][]{hoto13,ciol17}.
(b) For $\alpha_{\rm I}=40/9$, we use the same $L_{\rm preMAD}$, $\eta_{\bullet, \rm X}/f_{\bullet, \rm b}$ and $f_{\rm acc}$ as case (a),
except for $\tilde{t}_{\rm MAD}=8\times10^4$~s. This different choice of $\tilde{t}_{\rm MAD}$ only affects the obtained value of $M_{\rm fb}$ according to Equation (\ref{Mfb}),
and we get $M_{\rm fb}\simeq0.1M_{\sun}$. This value is also compatible with the maximum fall-back mass obtained from numerical simulations \citep[e.g.,][]{hoto13,ciol17}.
We note that the derived $M_{\rm fb}$ has a strong dependence on $\eta_{\bullet,\rm X}/f_{\rm b}$, $a_{\bullet}$ and
$\tilde{t}_{\rm PL}$ according to Equation (\ref{Mfb}). Smaller $M_{\rm fb}$ can be obtained if we adopt larger values of these parameters.

\begin{figure}
\includegraphics[width=\columnwidth]{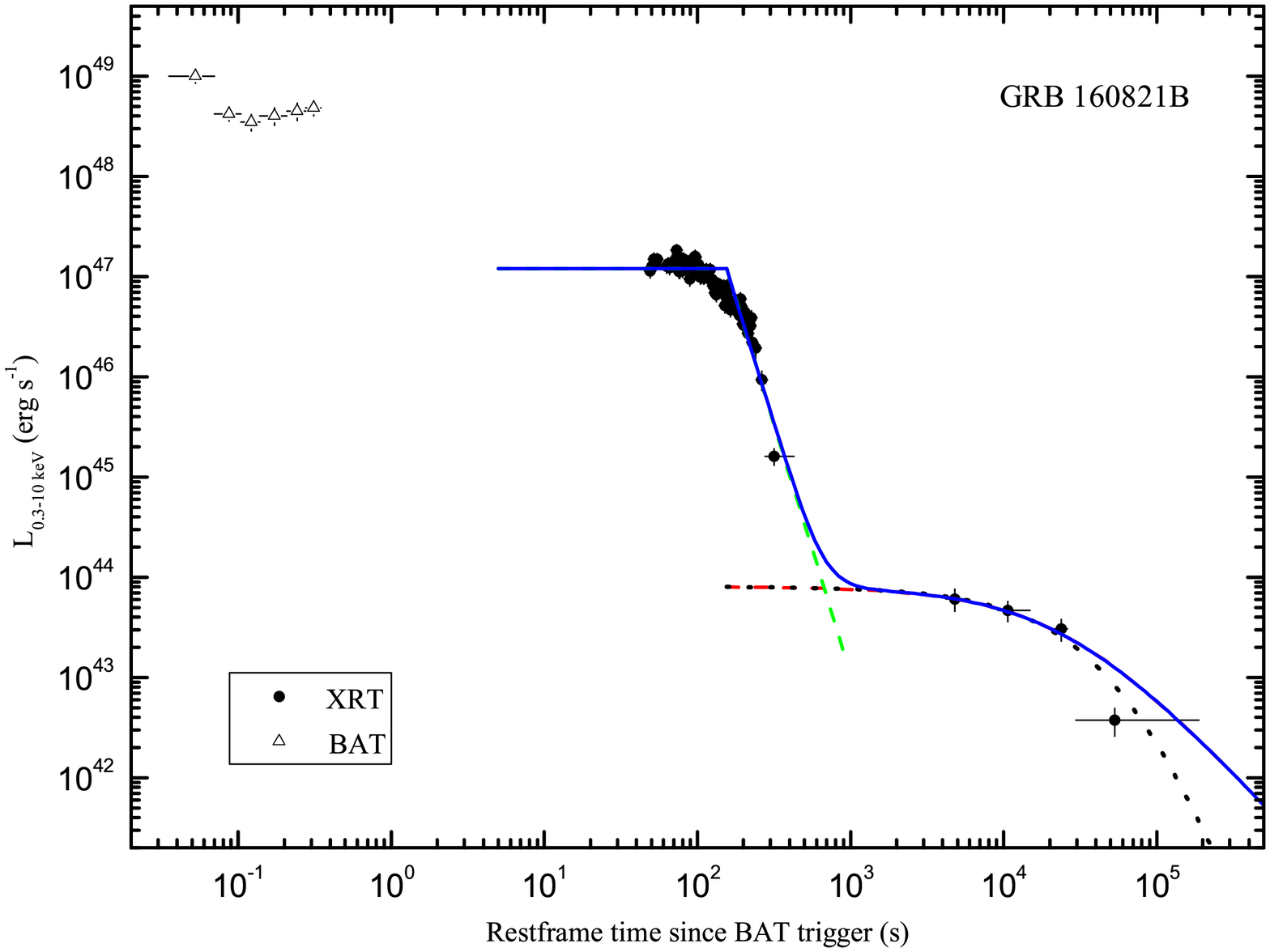}
\caption{Theoretical light curves as compared with the XRT data of GRB 160821B. The BAT and XRT data are exhibited with empty triangles and filled circles, respectively.
The dashed lines are our modeling results for the supra-massive magnetar (green) and the new-born BH (red, for the case of $\alpha_{\rm I}=5/3$),
and the blue solid line is the superposition of both components.
The adopted parameters are: $P_0=4$~ms, $B_{\rm p}=1.0\times10^{15}$~G, $M_\bullet=M_{\rm p}=2.37 M_{\sun}$, $a_\bullet=0.1$, $B_{\bullet,\rm MAD}=3.1\times10^{12}$~G,
$M_{\rm fb}=0.02M_{\sun}$, $\eta_{\rm X}/f_{\rm b}=\eta_{\bullet,\rm X}/f_{\bullet,\rm b}=1$ and $f_{\rm acc}=0.5$.
The case for $\alpha_{\rm I}=40/9$ is also shown with black dotted line, but only the BH component is exhibited for clarity.
The adopted parameters are the same as given above except for $M_{\rm fb}=0.1M_{\sun}$.
\label{160821B}}
\end{figure}

To compare our model with the data, we use Equations (\ref{L0}), (\ref{L_int}), and (\ref{L_lp})-(\ref{Lxlate}).
According to the fitting results of \citet{lv17}, we assume the luminosity declines with a slope of 5 after the collapse, and model the sharp decay phase
with $L_{\rm X,iso}=L_{\rm int}(t/t_{\rm b,int})^{-5}$.

Figure \ref{160821B} compares our theoretical 0.3--10 keV light curve with the XRT data.
The blue solid line is for $\alpha_{\rm I}=5/3$, while the black dotted line is for $\alpha_{\rm I}=40/9$. For clarity, the latter exhibits only the radiation component
produced by the BH. It is shown that
our model can describe the luminosity evolution rather well. However, our magnetar model does not explain the smooth transition from
the internal plateau to the sharp decay (at around $t_{\rm b,int}$). Since this transition happens during the magnetar collapse,
the data suggest that this process does not result in an abrupt cessation of emission in the X-ray band.
Meanwhile, the flux declines along with the spectral evolution \citep{lv17}, which
may be related to the smooth break in the light curve. The sharp decay phase may
be a joint result of the `curvature effect' \citep[e.g.,][]{feni96,kuma00,derm04} and the spectral evolution \citep{zhan09}.

\subsubsection{GRB 101219A}
GRB 101219A triggered the {\it Swift}/BAT  at 02:31:29 UT on 2010 December 19 \citep{gelb10} and was also detected by {\it Konus-Wind} \citep{gole10}.
The $\gamma$-ray  light curve shows a double-peaked structure with $T_{90}=0.6\pm0.2$~s \citep[15--150 keV;][]{krim10}. The time-integrated spectrum
is best fit in the 20~keV--10~MeV range by a PL with exponential cutoff model, which gives $E_{\rm pk}=490^{+103}_{-79}$~keV and a fluence of
$(3.6\pm0.5)\times10^{-6}$~erg~cm$^{-2}$ \citep{gole10}. With a redshift of $z=0.718$ \citep{chor10},
the resulting isotropic $\gamma$-ray energy in the observed 20~keV--10~MeV range is $E_{\gamma, \rm iso}\approx4.8\times10^{51}$ erg \citep{fong13}.

\begin{figure}
\includegraphics[width=\columnwidth]{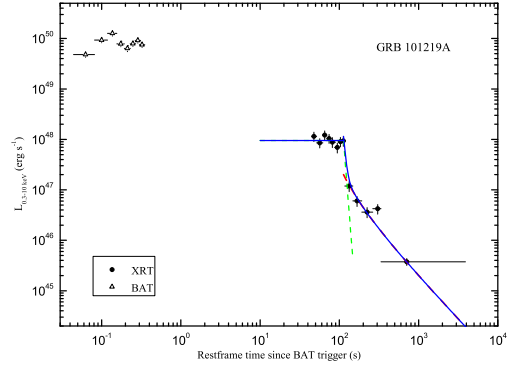}
\caption{Theoretical light curves as compared with the XRT data of GRB 101219A. The BAT and XRT data are exhibited with empty triangles and filled circles, respectively.
The dashed lines are our modeling results for the supra-massive magnetar (green) and the new-born BH (red),
and the blue solid line is the superposition of both components.
The adopted parameters are: $P_0=4$~ms, $B_{\rm p}=2.8\times10^{15}$~G, $M_\bullet=M_{\rm p}=2.37 M_{\sun}$, $a_\bullet=0.1$, $B_{\bullet,\rm MAD}=1.6\times10^{14}$~G,
$M_{\rm fb}=1.9\times 10^{-3}M_{\sun}$, $\eta_{\rm X}/f_{\rm b}=\eta_{\bullet,\rm X}/f_{\bullet,\rm b}=1$ and $f_{\rm acc}=0.5$.
\label{101219A}}
\end{figure}

The XRT began observing the field 60.5 s after the BAT trigger \citep{gole10}. The X-ray spectrum is best fit by an absorbed PL with $\Gamma_{\rm X}=1.8\pm0.1$
and $N_{\rm H}=6.6^{+2.3}_{-1.8}\times10^{21}$~cm$^{-2}$ \citep{fong13}. The light curve exhibits a short plateau before $\sim200$~s, then drop sharply, followed
by a single PL decay with slope of $1.91\pm0.08$ \citep{evan09}.
The {\it Swift}/UVOT commenced observations 67~s after the BAT trigger. No optical afterglow was detected within the  XRT position to a $3\sigma$ limit
of $\gtrsim 21.4$~mag in the {\it white} filter \citep{kuin10,fong13}. Observations by several ground-based instruments also revealed
no optical/NIR counterpart within the XRT error circle, and only $3\sigma$ upper limits were given \citep[e.g.,][]{pand10,covi10,fong13}.

The parameters of GRB 101219A can be estimated following the same way as GRB 160821B. For this burst,
we also assume $\eta_{\rm X}/f_{\rm b}=\eta_{\bullet,\rm X}/f_{\bullet,\rm b}=1$ and $f_{\rm acc}=0.5$.
By adopting $L_{\rm int}=9.5\times10^{47}$~erg~s$^{-1}$ and $t_{\rm b,int}=113$~s for the luminosity and duration of the internal plateau,
we obtain the upper limits of $P_0$ and $B_{\rm p}$: $P_0\lesssim 25$~ms and $B_{\rm p}\lesssim 1.1\times10^{17}$~G. Here we also adopt $P_0=4$~ms, then
the corresponding magnetic field strength and spin-down timescale are $B_{\rm p}\simeq 2.8\times10^{15}$~G and $\tau\simeq 4.6\times10^3$~s, respectively.
Using Equation (\ref{tcol}), we obtain the mass of the supra-massive magnetar $M_{\rm p}\simeq2.37M_{\sun}$.

Then the mass and spin of the new-born BH are $M_{\bullet}\simeq 2.37M_{\sun}$ and $a_{\bullet}\simeq0.1$, respectively. To obtain the values of $L_{\rm preMAD}$
and $\tilde{t}_{\rm MAD}$, we fit the data with Equation (\ref{Lxlate})
and get $L_{\rm preMAD}\simeq 2\times 10^{47}~{\rm erg}~{\rm s}^{-1}$ and $\tilde{t}_{\rm MAD}\simeq60$~s.
By substituting these values into Equations (\ref{B_preMAD}) and (\ref{Mfb}),
we obtain $B_{\bullet, {\rm MAD}}\simeq 1.6\times 10^{14}$~G and
$M_{\rm fb}\simeq1.9\times10^{-3}M_{\sun}$. The modeling results for the XRT data are shown
in Figure \ref{101219A}. For the magnetar radiation component, we have adopted a decay slope of 20 after the collapse according to the fitting results of \citet{lv15}.
It is shown that our model explains the afterglow data very well.

\subsection{Further Model Test}
\begin{figure}
\includegraphics[width=0.85\columnwidth]{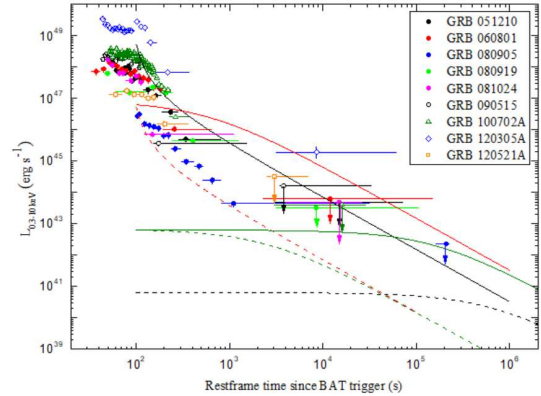}
\caption{Compare our model with the rest GRB light curves in our sample. The data as shown are XRT luminosity light curves calculated in the observed 0.3--10 keV energy band. The lines are our theoretical light curves calculated with the following parameters: black solid line:
$M_{\rm fb}=0.01~M_{\sun}$ and $B_{\bullet,\rm MAD}=10^{15}$~G; black dashed line: $M_{\rm fb}=0.01~M_{\sun}$ and $B_{\bullet,\rm MAD}=10^{11}$~G;
red solid line: $B_{\bullet, \rm MAD}=10^{14}$~G and $M_{\rm fb}=0.1~M_{\sun}$; red dashed line: $B_{\bullet, \rm MAD}=10^{14}$~G and $M_{\rm fb}=10^{-4}~M_{\sun}$; green solid line: $B_{\bullet, \rm MAD}=10^{12}$~G and $M_{\rm fb}=0.1~M_{\sun}$; green dashed line: $B_{\bullet, \rm MAD}=10^{12}$~G and $M_{\rm fb}=10^{-4}~M_{\sun}$. For other parameters used, see the text for details.
\label{xrt_lcs}}
\end{figure}

In Subsection \ref{search}, we have found a sample of 12 short GRBs with an internal plateau. Three of them have high-quality late X-ray data (GRB 070724A, GRB 101219A and GRB 160821B) while others show only one data point (GRB 120305A) or upper limits at $\sim 10^3-10^5$~s. The XRT light curves of the rest 9 GRBs with poor-quality late-time data are shown in Figure \ref{xrt_lcs}. For those without redshift measurements, an average redshift of 0.63 is assumed \citep{berg14}. We note that, except GRB 120305A, all other bursts show no evidence of an extra component
emerging after the sharp decay phase. However, even these upper limits are important for the consistency check of our model. To be self-consistent, the allowed parameter space of this model should be large enough to be compatible with these upper limits.

The most important parameters of our model are $M_{\rm fb}$ and $B_{\bullet, \rm MAD}$, which, however, are highly uncertain. To compare the theoretical light curves with the data, we use a range of values of these two parameters while leave other parameters fixed. Specifically, we
adopt $M_{\rm fb}=10^{-4}-10^{-1}~M_{\sun}$, $B_{\bullet, \rm MAD}=10^{11}-10^{15}$~G, and fix $M_{\bullet}=2.37~M_{\sun}$, $a_{\bullet}=0.1$,
$\eta_{\bullet,\rm X}/f_{\bullet,\rm b}=1$ and $f_{\rm acc}=0.5$. We also assume a rest-frame collapse time $t_{\rm col}/(1+z)=100$~s. Using these
parameters, we can calculate the theoretical light curves based on Equations (\ref{L_lp}), (\ref{t_blp}) and (\ref{Lxlate}) and an assumed redshift of 0.63.

To show clearly in Figure \ref{xrt_lcs}, we simply consider three cases: (1) $M_{\rm fb}=0.01~M_{\sun}$, $B_{\bullet,\rm MAD}=10^{11}-10^{15}$~G. The allowed luminosity regions are shown by two black boundary lines. We note that, as $B_{\bullet,\rm MAD}$ increases, the light curve gradually changes from type I to type II; (2) $B_{\bullet, \rm MAD}=10^{14}$~G, $M_{\rm fb}=10^{-4}-10^{-1}~M_{\sun}$. The corresponding luminosity regions are shown by two red boundary lines. In this case, the pre-MAD magnetic field strength is high, leading to type II light curves, and more fall-back mass produces higher luminosity; (3) $B_{\bullet, \rm MAD}=10^{12}$~G, $M_{\rm fb}=10^{-4}-10^{-1}~M_{\sun}$. The allowed luminosity regions are exhibited by two green boundary lines. This case results in type I light curves. Since $\tilde{t}_{\rm MAD}\propto M_{\rm fb}^{3/5}$ while $L_{\rm preMAD}$ has no dependence on $M_{\rm fb}$, more fall-back mass leads to longer plateau duration while leave the plateau luminosity constant. As shown in Figure \ref{xrt_lcs}, each case allows a large parameter space to be compatible with the data. It can also explain
the late excess of GRB 120305A by simply adjusting some parameters (see the red solid line). Therefore, our model is further supported by the rest GRBs in our sample.

\section{conclusion and discussion} \label{conclu}
Internal plateaus in GRB afterglows are commonly interpreted as the magnetic dipole emission from a supra-massive magnetar,
and the sharp decay may imply the collapse of the magnetar to a BH. Fall-back accretion onto the new-born BH can produce long-lasting activities
via the BZ process.
The magnetic flux accumulated near the BH would be confined by the accretion disks for a period of time, resulting in roughly a constant BZ luminosity.
As the accretion rate decreases, the magnetic flux is strong enough to obstruct gas infall and the MAD achieves. Then the BZ luminosity is determined by the
instantaneous accretion rate \citep[e.g.,][]{tche15}.
In the case of NS-NS mergers, we show that the BZ process could produce two types of typical X-ray light curves:
type I shows a long-lasting late plateau, followed by a decay with slopes ranging from 5/3 to 40/9;
type II exhibits roughly a single PL decay with slope of 5/3. The light curve patterns are mostly determined
by the magnetic field strength in the pre-MAD regime,
and are weakly affected by the initial mass accretion rate or the total fall-back mass.
Type I light curve requires low pre-MAD magnetic field strength,
say, $B_{\bullet, \rm MAD}\sim 10^{12}$~G,  while type II corresponds to relatively high $B_{\bullet, \rm MAD}$ values, say,
$B_{\bullet, \rm MAD}\sim 10^{14}$~G for typical parameters.
We search for such signatures of the new-born BH from a sample of short GRBs with an internal plateau,
and find two candidates: GRB 101219A and GRB 160821B, corresponding to
type II and type I light curve, respectively. By comparing the theoretical light curves with their XRT data,
we find our model can explain the data very well.
The derived total fall-back mass $M_{\rm fb}\sim 10^{-3}-10^{-2}~M_{\sun}$ is consistent with the results obtained from
numerical simulations. For the rest bursts with poor-qulity late X-ray data, our model are also compatible with observations.

Though the light curves of short GRBs with an internal plateau seem to support this scenario,
the sample is too small, and more observations are needed to
establish whether all such GRBs show light curves that are consistent with our model predictions.
The first question is how to distinguish between this internal emission produced by the BZ process with the external afterglow component. Maybe the easiest way is to diagnose from the light curve. The external afterglow typically shows a PL decay with slope of $\sim 1$, while the internal component exhibits either a long-lasting late plateau or a PL decay with slope of 5/3 or steeper. The late plateau component is more interesting. If the scenario of a collapsing supra-massive magnetar is preferred by an enlarged internal plateau sample in the future,
the simultaneous observation of a late X-ray plateau (e.g., GRB 160821B) would be a further support for this framework. Besides, multi-band afterglow observations could serve as an auxiliary diagnosis since the standard afterglow models have definite predictions to the afterglow evolutions.
Finally, we emphasize that the BH emission could, in principle, not be seen due to the following two reasons: first,
this emission could be hidden by the external afterglow component; second,
if the fall-back process is inefficient, the BH emission could be intrinsically weak and below the detection limit.
For the latter, as stated in Section \ref{model}, the magnetar accretion and propeller processes could greatly decrease the fall-back mass.
Outflows from the accretion disk could also interact with the fall-back material and reduce the BH emission \citep{fern15}.

It should be noted that the luminosity evolution in the MAD state and the value of $M_{\rm fb}$ are strongly
dependent on the mass accretion rate, which was assumed to be a fraction of the fall-back accretion rate in this work.
For example, if we assume $\dot M$ scales with $\dot M\propto t^{-4/3}$, the decay slope of the MAD luminosity would be different from 5/3,
and the value of $M_{\rm fb}$ would be an order of magnitude smaller than the results obtained in Section \ref{obs} \citep{kisa17}.
Besides, we did not consider the BH evolution during the accretion and BZ processes, which can in principle affect the BH mass $M_{\bullet}$ and spin parameter $a_{\bullet}$ \citep[e.g.,][]{chen17,lei17}. However, for very low accretion rate as studied in this work, this effect might be ignored. From another point of view, some sacrifice in accuracy may be justified, given that some related physical processes (e.g., the accretion and propeller processes, the GW effect and the interaction between outflows and fall-back matter) were not taken into account in our model.
Finally, it is interesting to further investigate why type I and type II light curves correspond to much different pre-MAD magnetic field strength, which is beyond the scope of this work.

\citet{kisa15} employed the same BZ process to interpret the extended emission of short GRBs and their X-ray afterglows.
Considering that the internal plateau and extended emission have similar durations and continuous luminosity distribution \citep{lv15},
their model can also explain the light curve of GRB 160821B-like bursts \citep{kisa17}. We emphasize that our model is different from theirs on three points:
(1) different central engines. We assume the product of the NS-NS merger is a supra-massive magnetar which collapses to a BH at late times, while their assumed
central engine is a prompt BH; (2) different theoretically light curves. Our model predicts two types of typical X-ray light curves, while theirs can only
produce the one with a plateau followed by a steep decay, corresponding to our type I light curve. It is easy to understand. Our type II light curve
is intrinsically due to the zero point effect, which changes the zero point from the beginning time of the BH accretion to the burst trigger time.
While there is no such transformation in the case of a prompt BH central engine; (3) on the maximum decay slope after the internal plateau. Their model predicts a maximum slope of 40/9, while ours can produce a much steeper decay. Some internal plateaus followed by a decay with slope as steep as $\sim 10$ seem to support the magnetar collapse scenario \citep[e.g.,][]{rowl13,lv15}.

The two scenarios can, in principle, be distinguished by observations.
As the supra-massive magnetar collapses, the magnetic field  would be ejected as the event horizon swallows the star based on the `no-hair theorem'.
The entire magnetic field outside the horizon detaches and reconnects, resulting in intense
electromagnetic emission in a short time \citep{baum03,lehn12,dion13}. One such product is a bright radio `blitzar', which was proposed as a likely
source of fast radio bursts \citep[FRB;][]{falc14}. This FRB-like event, if observed at the end of the internal plateau, can be an evidence of the
collpse of a supra-massive magnetar to a BH \citep{zhan14}.

Our model has two important implications:

First, considering the similarity of the internal plateau and extended emission of short GRBs, our model may have the potential to explain
the extended emission and the X-ray afterglows. Within the magnetar scenario, the spin-down process with or without a significant accretion
was investigated to explain the extended emission \citep{metz08b,bucc12,gomp14}. The magnetar might collapse at some time, then the X-ray afterglow
could be interpreted as the radiation from the BZ process of the new-born BH. Interestingly, most of the afterglow light curves can be best fit
by a single PL \citep{lv15}, while only a minority show a long-lasting plateau (e.g., GRB 060614). These features seem to resemble our theoretical
light curves. This issue will be studied in future work.

Second, as suggested by \citet{kisa15}, the long-lasting activities of the new-born BH would significantly affect the macronovae.
Macronovae could be powered by the radioactivity of r-process elements synthesized in the ejecta of a NS-NS merger \citep[e.g.,][]{li98,kulk05,metz10,barn13},
or by the energy injection from the central engine, e.g.,a BH or a stable magnetar \citep[e.g.,][]{yu13,metz14,gao15,kisa_etal15,kisa16}.
Recently, \citet{kisa16} proposed
a X-ray powered model in which the X-ray excess \citep[e.g., GRB 130603B;][]{fong14} gives rise to the simultaneously observed infrared excess via thermal re-emission.
However, their model did not specify the mechanism of the X-ray excess.
Our model provides a possible mechanism for such kind of X-ray excess, and the X-ray powered macronovae will be further studied in a separated paper.

In our model, the late X-ray afterglows of short GRBs with an internal plateau are produced by the BZ process of a new-born Kerr BH, the magnetic field of which is supported by the surrounding disk. Recently, \citet{nath17} showed that the collapse of a rotating magnetized NS would
leave behind a charged spinning (Kerr-Newman) BH. Such a charged BH was also proposed by \citet{zhangb16} as a product of BH-BH mergers, of which at least one carries a certain amount of charge \citep[see also][]{lieb16}. In our study, if the product of collapsing supra-massive NS is a Kerr-Newman BH, then the BZ power can be provided by the BH itself even if there is no fall-back accretion. A quantitative comparison between this model and ours is interesting, which is beyond the scope of this work.

\section*{Acknowledgements}
We acknowledge the anonymous referee for helpful comments and suggestions. We also thank Bing Zhang and He Gao for helpful discussions.
This work made use of data supplied by the UK Swift Science Data Centre at the University of Leicester.
This study was supported by the Strategic Priority Research Program of the Chinese Academy of Sciences (Grant No. XDB23040400).
S. L. Xiong was also supported by the Hundred Talents Program of the Chinese Academy of Sciences (Grant No. Y629113).
W. H. Lei and W. Chen acknowledge support from the National Natural Science Foundation of China (Grant U1431124).
B. B. Zhang  acknowledges support from the Spanish Ministry Projects AYA2012-39727-C03-01 and AYA201571718-R.
L. M. Song acknowledges support from the National Program on Key Research and Development Project (Grant No. 2016YFA0400801)
and the National Basic Research Program of China (Grant No. 2014CB845802).

\bsp	
\label{lastpage}
\end{document}